# The Theory of SERS on Dielectrics and Semiconductors

V.P. Chelibanov[1] A.M. Polubotko[2]

[1]State University of Information Technologies, Mechanics and Optics, Kronverkskii 49, 197101 Saint Petersburg, RUSSIA  E-mail: Chelibanov@gmail.com

[2] A.F. Ioffe Physico-Technical Institute, Politechnicheskaya 26, 194021 Saint Petersburg, RUSSIA E-mail: alex.marina@mail.ioffe.ru

## Abstract

It is demonstrated that the reason of SERS on dielectric and semiconductor substrates is the enhancement of the electric field in the regions of the tops of the surface roughness with very small radius, or a very large curvature. The enhancement depends on the dielectric constant of the substrate and is stronger for a larger dielectric constant. It is indicated that the enhancement on dielectrics and semiconductors is weaker than on metals with the same modulus of the dielectric constant.  The result obtained is confirmed by experimental data on the enhancement coefficients obtained for various semiconductor and dielectric substrates.

## Introduction

Surface Enhanced Raman Scattering (SERS) was discovered in 1974 [1] and is a well known effect at present. First it was observed on rough surfaces of silver. Further SERS was found on gold and copper [2-4]. These results were the reason that it was considered that the reason of SERS are so-called surface plasmons, or some excitations of an electron gas of a metal, which result in enhancement of an electromagnetic field near the surface. The most enhancement in these experiments was of the order $\sim 10^6$. However, further SERS was discovered on dielectric and semiconductor substrates. The most important feature of SERS on these substrates is a lower enhancement coefficient, which achieves the value $\sim 10^4$. Its maximum value is $\sim 10^5$. The most full review on SERS on semiconductor and dielectric structures one can find in [5, 6]. Numerous investigations demonstrate that the enhancement coefficient for the molecules, adsorbed in the first layer, which interact with the substrate directly is more than the one for the molecules in the second layer something about $\sim 10^2$. In addition the enhancement is observed in higher layers [7], however it disappears for the heights, which are more than some value. This behavior allows to conclude that there is a short range enhancement, which is named as a "chemical" enhancement associated with a direct interaction of the molecules with the substrate and a long range enhancement, associated with an electromagnetic field. Just the last enhancement one associates with existence of so-called surface plasmons. However observation of SERS on dielectrics and semiconductors evidently demonstrates that the plasmon mechanism is absent in this case. Therefore one usually associates the enhancement on dielectrics and semiconductors with the "chemical" mechanism. This means that the enhancement is caused by the distortion of the electron structure and polarizability of the malecule due to adsorption. The last assertion can be rejected by the following reasoning. The "chemical mechanism" assumes that the cross-section increases due to the direct interaction of the molecule with the substrate. However it is obvious, that the distortion of the electron structure can result in both the increase and decrease of the cross-section since the fact of the distortion is associated with the change of the totality of matrix elements, which determine the expression of the cross-section. In addition this effect must be present not only on a rough but on single surfaces. However in accordance with the results of [8, 9], the increase of the Raman cross-section on the single surfaces is absent. In addition we demonstrated in [10, 11] that the predominant enhancement of Raman scattering in the first layer of adsorbed molecules, or the "chemical mechanism" is a pure electrodynamical effect, associated with a very strong change of the electric field near the tops of the roughness, when one moves away from the surface. These sharp points are named as active sites or hot spots. The electric field and its derivatives differ so strongly in space that this behavior explains the "chemical effect" completely. Thus these results point out the incorrectness of the ideas of the plasmon and the "chemical" mechanisms.

## The SERS theory on the semiconductor and dielectric Substrates

As it is well known the reason of SERS is the presence of surface roughness both on metal and on dielectric and semiconductor substrates. The electromagnetic field in the region of the roughness strongly differs from the field in a free space. This follows from the fact that a free space possesses by rotational and translational symmetry. Therefore the electromagnetic field has the form of plane waves in a far region with the characteristic size, which is equal to the wavelength $\lambda$. However in the region of the surface roughness near the surface the space is strongly inhomogeneous. Therefore the field must change strongly with a characteristic size $l_E$, which is equal to the characteristic size of the roughness. Therefore there is a so-called surface or a near field near the surface. Mathematically its appearance is associated with the necessity to satisfy boundary conditions for the electric and magnetic fields $\mathbf{E}$ and $\mathbf{H}$. The existence of this field is well known from radiophysics and electrodynamics. It presents always and its nature is associated just with irregular character of the surface. It strongly decreases when one moves away from the surface and therefore it is named as a surface, or a so-called near field, as it is accepted in electrodynamics. Its nature does not associated with excitation of the electron plasma of the metal. It exists not only near metal, but near the rough surfaces of semiconductors and dielectrics. It is necessary to note specially that this fact is ignored completely in the works on SERS and is substituted by ideas of existence of the surface plasmons.

Let us consider the main properties of the surface field near a model of a rough surface, the lattice of an eshelett type (Fig. 1). Its consideration does not differ from the ones in [12, 13].

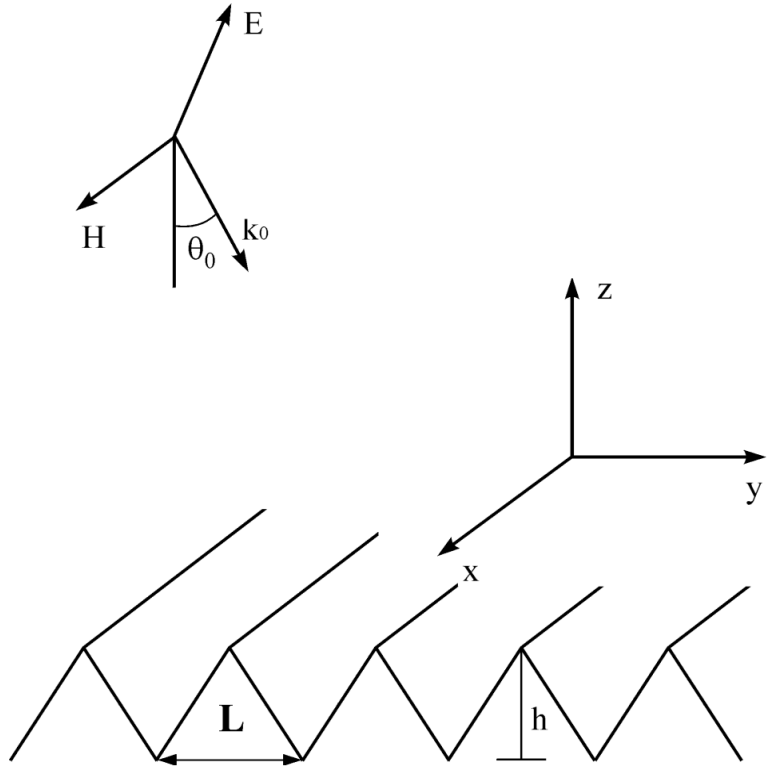

Figure 1. Diffraction of a plane H - polarized electromagnetic wave on the lattice of the eshellet type. $L$ - is the lattice period, $h$ - is the height of the lattice. The material of the lattice can be both the metal and dielectric or semiconductor type.

The only feature is that we shall consider and compare the cases of a metal and dielectric lattices. The electromagnetic field can be represented in the form

$$\mathbf{E} = \mathbf{E}_{inc} + \mathbf{E}_{surf..sc},$$

where

$$\mathbf{E}_{inc} = \mathbf{E}_{0,inc} e^{-ik_0 \cos\theta_0 z + ik_0 \sin\theta_0 y}$$
$$\left|\mathbf{E}_{0,inc}\right| = 1$$

,

$\theta_0$ -- is the incident angle, $k_0 = \left|\mathbf{k}_0\right|$ - is a modulus of the wave vector of the incident field in a free space

$$\mathbf{E}_{surf.sc.} = \sum_{n=-\infty}^{+\infty} \mathbf{g}_n e^{i\alpha_n y + i\gamma_n z} , \qquad (1)$$

$$\gamma_n = \sqrt{k_0^2 - \alpha_n^2} ,$$

$\mathbf{g}_n$ - are the vectors and $n$ - the numbers of spacial harmonics. For the lattice with a period $L \ll \lambda$ (here $\lambda$ is the wavelength of the incident light), the spatial harmonic with $n = 0$ is a direct reflected wave, while all others are inhomogeneous plane waves, which are strongly localized near the surface. The maximum localization size has the harmonic with the number $n = 1$. All the others are localized significantly strongly. Precise solution of the diffraction problem reduces to determination of the $\mathbf{g}_n$ values. The main feature of the surface field is existence of a singularity of the electric field near the wedges of the surface or so-called rod effect. This type of the behavior does not depend on the specific profile of the lattice and is determined by existence of the sharp wedges. In addition this behavior does not depend on the dielectric properties of the lattice and exists on lattices with any dielectric constant, which differs from the dielectric constant of vacuum

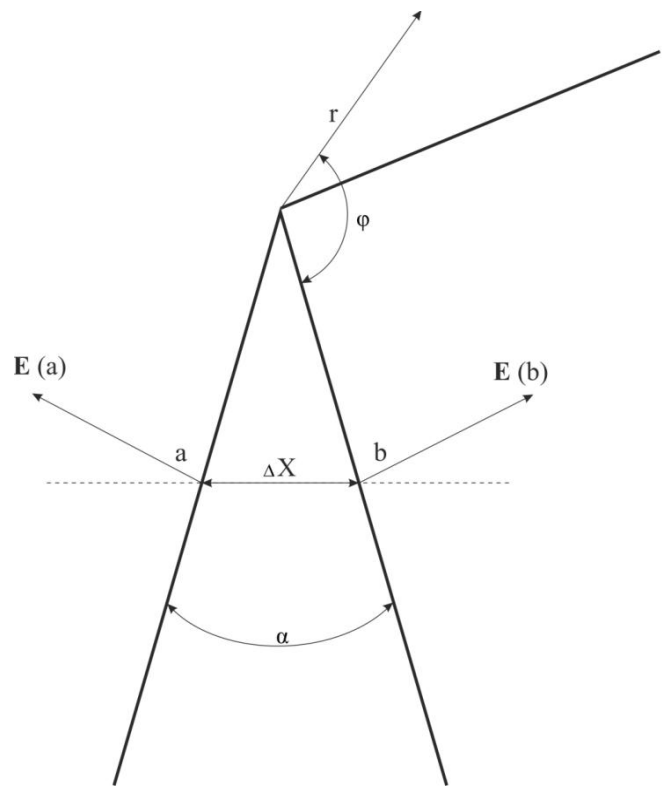

Figure 2. The wedge. Here we point out the behavior of the electric field near an ideally conductive wedge, where it is perpendicular to the surface.

For the metal ideally conductive lattice and for $H$ polarized wave, which is characterized by the component of the magnetic field $H_x$, near the wedge (Fig. 2) the electric field can be estimated as

$$E_r \approx -|\mathbf{E}_{o,inc}| C_0 \left(\frac{l_1}{r}\right)^\beta \sin(\lambda_1 \varphi)$$
$$E_\phi \approx -|\mathbf{E}_{0,inc}| C_0 \left(\frac{l_1}{r}\right)^\beta \cos(\lambda_1 \varphi)$$
, (2)

where $C_0$ is some numerical coefficient, ($l_1 = L$ or $h$) is a characteristic size in the lattice

$$\lambda_1 = \pi/(2\pi - \alpha) \quad ,$$

$\alpha$ - is the wedge angle,

$$\beta = 1 - \lambda_1 = \frac{\pi - \alpha}{2\pi - \alpha} \quad .$$

The peculiarity of the field behavior (2) is appearance of the singularity $(l_1/r)^\beta$, which describes a pure geometrical nature of the field enhancement. It determines the following behavior of the $\mathbf{g}_n$ coefficients in expression (1)

$$\mathbf{g}_n \sim |n|^{\beta-1} \quad . \tag{3}$$

Indeed after substitution of (3) in (1) one can obtain

$$\sum_{\substack{n=-\infty \\ n \neq 0}}^{+\infty} |n|^{\beta-1} e^{-2\pi|n|z/L} \sim 2\int_0^\infty t^{\beta-1} e^{-2\pi t/L} dt \sim 2\left(\frac{L}{2\pi z}\right)^\beta \quad .$$

For the wedge angles in the interval $0 < \alpha < \pi$, the $\beta$ value for an ideally conductive lattice changes in the interval $0 < \beta < 1/2$, and the coefficients $\mathbf{g}_n$ slowly decrease with the increase of $n$. Thus the singular behavior of the field arises due to specific adding of the surface waves near the top of the wedge.

Let us consider the case of an non ideal conductive lattice, when it can be both of a metal and a dielectric type [13, 14]. Here we consider the case of the $H$ polarized wave, which is characterized by the $H_x$ component of the magnetic field. We shall try to find a singular solution for the electric field, which must dominate near the top of the dielectric wedge. Let us try to find the solution in the form

$$H_x^{(I,II)} = C_0^{(I,II)} r^\tau$$

Here the upper indices $I$ and $II$ designate two media: vacuum - $I$, and the wedge - $II$. $H_x^{(I,II)}$ should satisfy the equation

$$\Delta H_x^I + k_0^2 H_x^I = 0 \qquad (4)$$

in the first medium $I$ (in vacuum) and

$$\Delta H_x^{II} + k_0^2 \varepsilon H_x^{II} = 0 \qquad (5)$$

for the medium $II$ (the wedge). The last ones can be both of the dielectric and metal types. In the last case the metal can be characterized by a negative dielectric constant. The $E_r$ and $E_\varphi$ components in the medium with the dielectric constant $\varepsilon$ are determined as

$$E_r = -\frac{1}{i\omega\varepsilon} \frac{1}{r} \frac{\partial H_x}{\partial \varphi},$$

$$E_\varphi = \frac{1}{i\omega\varepsilon} \frac{\partial H_x}{\partial r}.$$

The solutions of the equations (4, 5) for $H_x^{(I,II)}$, and the boundary conditions of the continuity of the tangential components of the electric and magnetic fields on the wedge surface ($\varphi = 0$ and $\varphi = 2\pi - \alpha$ (Fig. 2), $\alpha$ is the wedge angle)

$$H_x^I = H_x^{II} \qquad (6)$$

$$E_r^I = E_r^{II} \qquad (7)$$

can be reduced to the equation for determinations of the coefficients $C_0^{(I,II)}$

$$\frac{d^2 C_0^{(I,II)}}{d\varphi^2} + \tau^2 C_0^{(I,II)} = 0.$$

The general form of its solution has the form

$$C_0^{(I,II)}(\varphi) = A^{(I,II)} \cos \tau\varphi + B^{(I,II)} \sin \tau\varphi.$$

After substitution of this solution in the boundary conditions (6, 7) one can obtain a transcendental equation for determination of $\tau$

$$F_1(\tau) + \frac{F_2(\tau)}{\varepsilon} + \frac{F_1(\tau)}{\varepsilon^2} = 0, \qquad (8)$$

where

$$F_1(\tau) = \sin(2\pi - \alpha)\tau \times \sin \alpha\tau,$$

$$F_2(\tau) = 2[1 - \cos(2\pi - \alpha)\tau \times \cos \alpha\tau].$$

One can see from equation (8) that for $\varepsilon \to \pm\infty$, that corresponds to a very large value of the dielectric constant of the dielectric, or an ideally conductive wedge has the form

$$F_1(\tau) = 0.$$

Since we need in a singular solution for $E_r$ and $E_\varphi$ then the corresponding value of $\tau$, for the electric field components, which satisfies to the condition of the finiteness of the energy in the region of the top of the wedge

$$\int\limits_{V \to 0} (|\overline{E}|^2 + |\overline{H}|^2) dV \neq \infty,$$

is equal to

$$\tau = \lambda_1 = \frac{\pi}{2\pi - \alpha} .$$

First of all it is necessary to note that $E_r$ and $E_\varphi$ are singular and $E_r, E_\varphi \to \infty$ when $r \to 0$ that results in the enhancement of optical processes both on the metal and dielectric wedge and to the enhancement in Raman scattering in particular. Since we study the enhancement on dielectric lattices, then we need to solve the equation (8) for $\tau$ for a positive value of the dielectric constant $\varepsilon$. One can do it by an asymptotic expansion of the equation (8) on $(1/\varepsilon)$. However the problem is in the convergence of this expansion for real values of the dielectric constant, which corresponds to semiconductor and dielectric substrates. Here we shall give approximate expression for $\tau$, which is valid only for sufficiently large values of the dielectric constant

$$\tau \cong \frac{\pi}{2\pi - \alpha} + \frac{\tau_1}{\varepsilon} + O\left(\frac{1}{\varepsilon}\right)^2 ,$$

where

$$\frac{\tau_1}{\varepsilon} = \frac{2\left[1 + \cos\frac{\alpha\pi}{2\pi - \alpha}\right]}{\varepsilon(2\pi - \alpha)\sin\frac{\alpha\pi}{2\pi - \alpha}} .$$

Let us consider this value for small angles of the wedge $\alpha$. Then the condition that $\frac{\tau_1}{\varepsilon}$ is small with respect to $\frac{\pi}{2\pi - \alpha}$ is

$$\varepsilon \gg \frac{8}{\pi\alpha} .$$

Then the value $\beta = 1 - \tau$, which determines the singular behavior of the components of the electric fields

$$E_r, E_\varphi \sim \left(\frac{l}{r}\right)^\beta, \tag{9}$$

increases with the increase of $\varepsilon$ that determines stronger enhancement. The result obtained demonstrates that for small values of the angle $\alpha$ one must increase the value of the modulus of the dielectric constant in order to obtain sufficiently strong enhancement. This result can be interpreted as follows. For the decrease of the angle $\alpha$ the wedge becomes more transparent for the electromagnetic field and this system becomes "less inhomogeneous", that is expressed in the decrease of the index $\beta$ in (9) and hence in the decrease of the enhancement of the components of the electric field $E_r$ and $E_\varphi$. In order to increase the enhancement it is necessary to "increase the inhomogeneity" of the system that can be made by the increase of the modulus of the dielectric constant.

One should note a very important property. The $\beta$ value, which determines the enhancement near the wedge is larger for a metal wedge with the same modulus of the dielectric constant compared with the dielectric wedge. This fact determines a more enhancement for a metal wedge and is associated with a more "inhomegeneity of the space". A metal expels the electric field that is the reason of the stronger inhomogeneity of such medium compared with a dielectric, which is transparent for the field.

For the values of the dielectric constant, which correspond to real semiconductors and dielectrics it is necessary to take into account the next terms of the expansion of $\tau$ in the row by the $(1/\varepsilon)$ degrees. Therefore in order to form more precise ideas about the enhancement near the dielectric wedge we performed a numerical solution of the equation (8) for determination of $\tau$ and of the index $\beta = 1 - \tau$ in (9) for various values of the dielectric constant and various values of the angle $\alpha$. The results for the index $\beta$ are presented on the figure 3. One can see that the more is $\varepsilon$, then the more is the value $\beta$ for the same values of $\alpha$. This fact points out the more is the "space heterogeneity" then the more is the index $\beta$ that results in the stronger enhancement. First the value $\beta$ increases with the increase of $\alpha$ and then it begin to decrease. The increase of $\beta$ is associated with the "enhancement of the inhomogeneity of the space" and the electric field penetrates in the dielectric with a lesser degree. When $\alpha \to 0$ the dielectric wedge disappears and the space becomes absolutely homogeneous. However when $\alpha$ achieves some values and becomes more, the index $\beta$ begins to decrease since the angle at the top of the wedge increases. For $\alpha \to \pi$ the singularity disappears and we have a plain boundary where the field behavior is described by the Fresnel refraction and reflection laws.

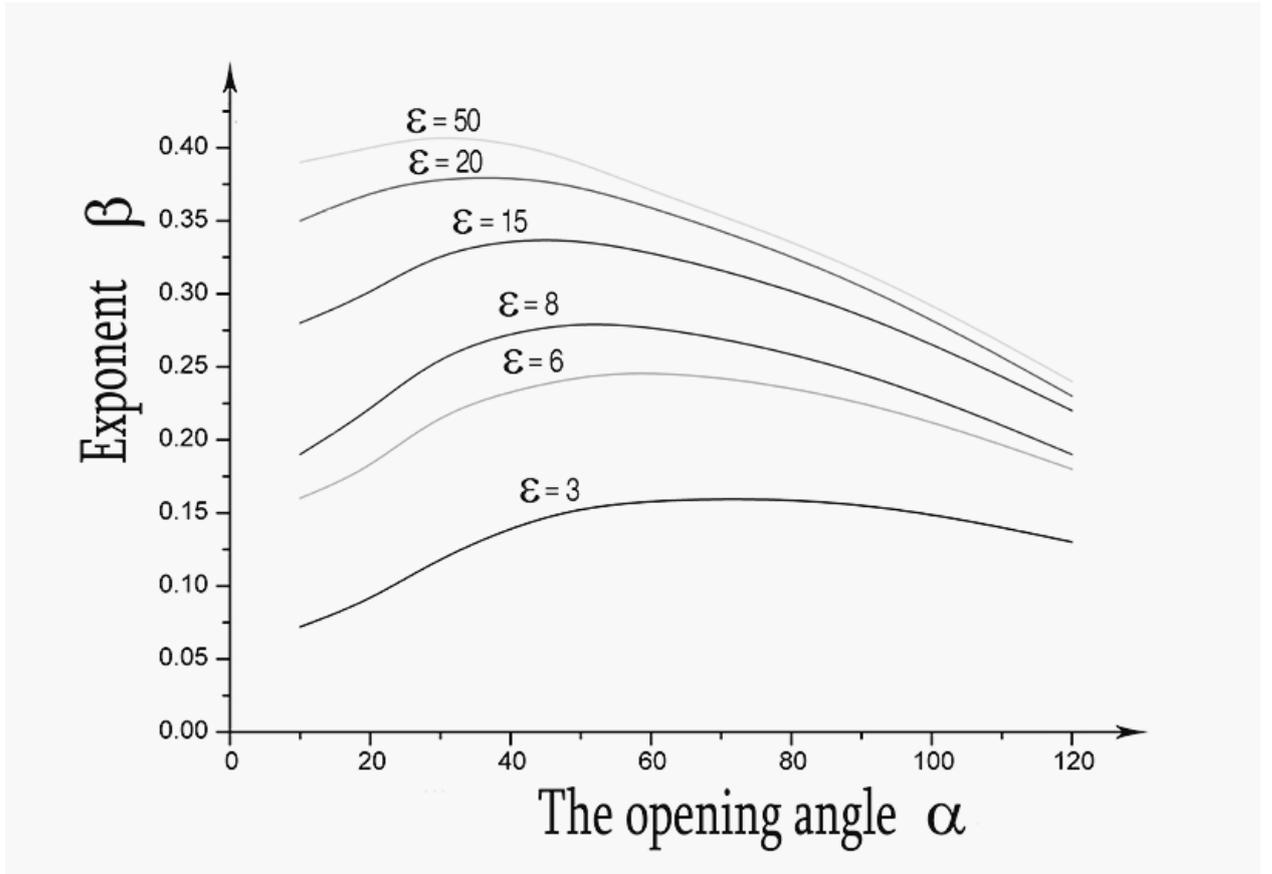

Figure 3. Dependence of the index $\beta$ on the wedge angle for various values of the dielectric constant of the wedge $\varepsilon$.

The model of the roughness of the wedge form is a typical case, when one can speak that the medium properties change strongly on very short distances. This property first of all refers to the area of the top of the wedge. In case we move along the dashed line (Fig. 2) first we have vacuum, then sharply pass to another medium, dielectric, with another value of the dielectric constant and then again pass into vacuum. In the region of the top of the wedge this crossing occurs on a very small distances that results in a very strong change of the field and to the appearance of a singularity of the electric field derivative near the top of the wedge.

$$\frac{\mathbf{E}(b) - \mathbf{E}(a)}{\Delta X} \to \infty \quad \text{при} \quad \Delta X \to 0$$

The requirement that the field tends to infinity near the top of the wedge can be explained by the fact that the direction of the electric field on the top of the field is not defined and we would have a case of not a physical result. Therefore the infinity value of the field in this point avoids this nonuniqueness. For a more real roughness in the form of a rounded wedge (Fig. 4) one can state that there is an enhancement of the electric field in the region of the maximum curvature of the top. In addition, in case of the dielectric wedge there can be the enhancement both of the normal and tangential component of the electric field. The above regularities, which describe the enhancement of the field in the areas of a maximum curvature and also lower enhancement on semiconductor and dielectric substrates must preserve for more realistic three dimensional roughness.

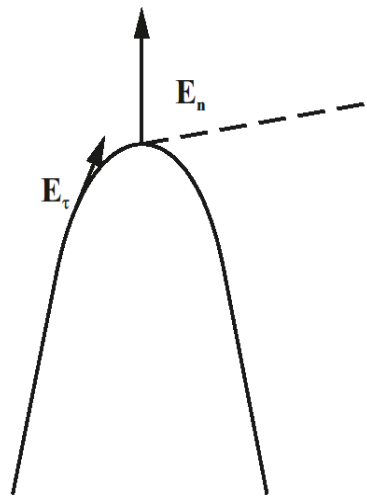

Figure 4. The model of the roughness of the rounded wedge. In case of semiconductor or dielectrics both the normal and the tangential components of the electric field can be enhanced.

It is necessary to note that any SERS experiments are performed on various nano particles (Fig. 5).

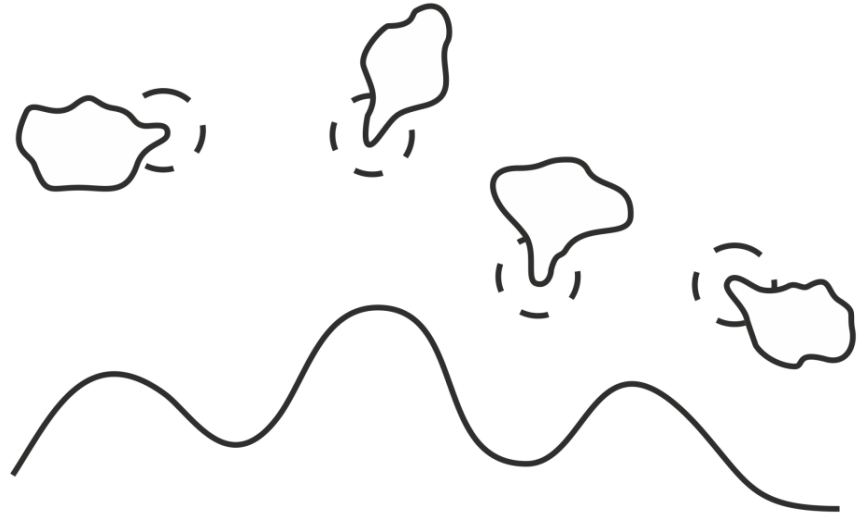

Figure 5. The system of colloidal particles. The enhancement arises in the areas of sharp points inside the dashed circle lines

The form of these particles is the matter of principle. The strong enhancement occurs on the particles with a sharp form, while spherical particles do not result in a strong enhancement. This result follows first of all from an analytical expression for the enhancement coefficient for the electric field near the sphere

$$G = \frac{\varepsilon - 1}{\varepsilon + 2} \tag{10}$$

For the case $|\varepsilon| \gg 1$, that usually is always valid for the wavelengths of the incident radiation, the enhancement is practically absent both for a metal and dielectric spheres, while for the roughness of the wedge or a tip form it tends to infinity. This result was confirmed in experimental investigations of Emory и Nie [15] who demonstrated that the enhancement arises on the particles with sharp points, while on spherical particles it is practically absent. As for the case of a metal the most sharp regions of the roughness on semiconductors and dielectrics are so-called "active sites" or "hot spots", where the enhancement is maximal. The fact that the enhancement on semiconductors and dielectrics is significantly lower than on metals is well confirmed by experimental results. For example the enhancement for the 4-mercaptopyridine molecule, adsorbed on CdS for the line 1016 см$^{-1}$ is $\sim 10^2$ [16], on ZnS $\sim 10^3$ [17], on ZnO $\sim 10^3$ [18], on CuO $\sim 10^2$ [19]. SERS also was observed on $TiO_2$ in [20-25]. The authors of [21] point out the enhancement factor $\sim 10^3$.

It is necessary to note also that SERS on semiconductors and dielectrics must possesses by the following regularities, which are well described in the monograph [7]. It is the first layer effect, when the enhancement from the first layer of adsorbed molecules is significantly stronger than the one from the second and upper layers. It is the frequency dependence of SERS, which must be a superposition of the $(\hbar\omega)^4$ law characteristic for a usual Raman scattering, and of the frequency dependence of the dielectric constant of the semiconductor or dielectric. The important issue is the role of the quadrupole light-molecule interaction in SERS on these substrates. Since the enhancement in this case is lower, than the electric field derivatives are significantly lower than those on the metal substrates. Therefore the forbidden lines in the SERS spectra of molecules with sufficiently sufficiently high symmetry on these substances will be significantly weaker or absent at all.

One should note that the enhancement both on metal and semiconductor or dielectric objects, when their characteristic sizes are significantly less than the wavelength $(l_1 << \lambda)$ is proportional to the size of these objects. It is a well known result, which is directly seen from the formula (2) for the electric field components. This result is the consequence of the electrostatic approximation, which is used for receiving of these formulae.

## Some notions concerning of interpretation of the SERS mechanism

It is necessary to note that there are some numerical results on the electric field strength, which were obtained by some authors, when they considered various objects with wedges and sharp points, [26] for example. As usual they obtained the most enhancement near the wedges or sharp points. However they assigned this enhancement to excitation of plasmons. One should note again that the enhancement is associated just with the sharp points and a sharp change of the dielectric constant on very small distances. Therefore this effect is mandatory, but its assignment to the plasmon excitation is deeply erroneous. It is well known that SERS is observed not only on rough surfaces but on molecules adsorbed on colloidal particles of metals, dielectrics and semiconductors, or on single nanoparticles. In particular many works point out appearance of the strong enhancement or appearance of "active sites", or "hot spots" in the areas between very closely situated nanoparticles.

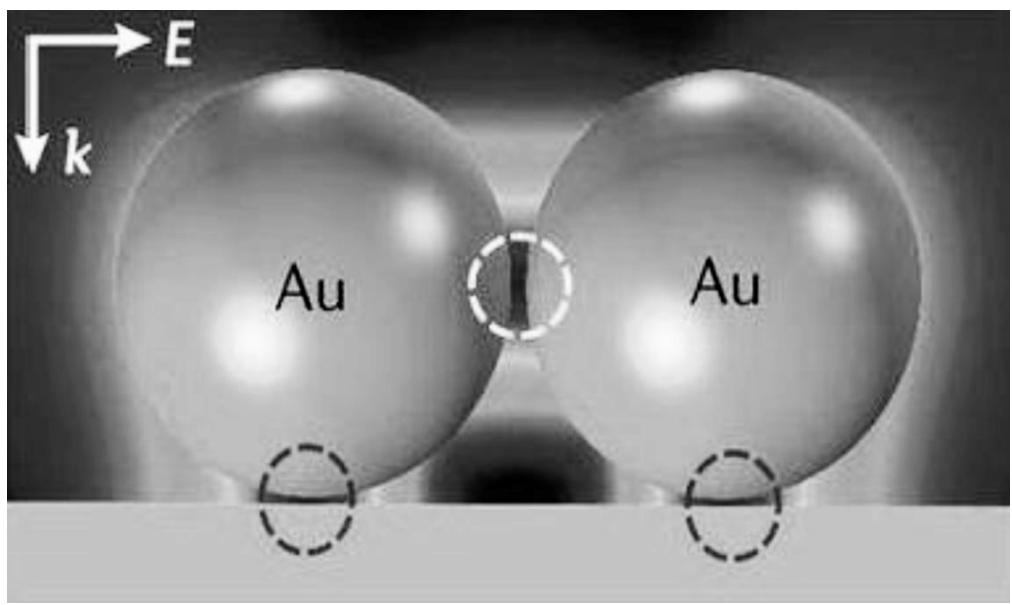

Figure 6. The model of a closely situated spherical nano particles. The most enhancement occurs in the small area between nanoparticles, where both and its derivatives change very strongly because of the necessity to satisfy the boundary conditions on the sphere surfaces.

Such situation usually is described with the help of two or more spherical particles (see on the areas, designated by dashed circles on figure 6). Usually the authors consider that the reason of this enhancement is excitation of surface plasmons in the areas between these particles, or between the nanoparticle and the substrate. One should note that there is a very strong change of the electric field and its derivatives in these regions because of the necessity to satisfy the boundary conditions. This effect is specially strong, when the spheres are nearly touched. Therefore from our point of view this fact results in the strong increase of the electric fields and its derivatives in these regions and is the reason of the enhancement both of the dipole and quadrupole light-molecule interactions.

## Acknowledgement

The authors want to thanks Dr. A.V. Golovin from Saint Petersburg State University for the performance of numerical calculations.